\documentclass[10pt,twocolumn,letterpaper]{article}

\usepackage{cvpr}
\usepackage{times}
\usepackage{epsfig}
\usepackage{graphicx}
\usepackage{amsmath}
\usepackage{amssymb}
\usepackage{appendix}
\usepackage{listings}
\usepackage{xcolor}

\usepackage{listings}
\usepackage{xcolor}

\lstdefinelanguage{Lean}{
    morekeywords={
        theorem, lemma, example, def, by,
        have, let, exact, apply, rw, intro,
        obtain, induction, cases, simp
    },
    sensitive=true,
    morecomment=[l]{--},
    morestring=[b]"
}

\usepackage[pagebackref=false,breaklinks=true,letterpaper=true,colorlinks,bookmarks=false]{hyperref}

\cvprfinalcopy 

\def\cvprPaperID{****} 

\ifcvprfinal\fi

\begin{document}

\title{Towards a Formalised Framework for Analog Electronics and Circuit Theory}

\author{Yaohua Zhang\\
University College London\\
{\tt\small yaohua.zhang.20@ucl.ac.uk}
\and
Zhi Kai Pong\\
Independent Researcher\\
{\tt\small zhikaip@gmail.com}
}

\maketitle

\begin{abstract}
Circuit theory provides the mathematical foundation for the analysis and design of electronic circuits. However, conventional circuit derivations often rely on assumptions and mathematical relationships that are not explicitly recorded, making their verification and reuse challenging. In this paper, we present the first step towards a formalised framework for circuit theory and analog electronics using the Lean 4 interactive theorem prover. As a proof of concept, we formalise an ideal RC low-pass filter, establishing reusable definitions for component impedance and circuit connections, and machine-checked theorems for voltage division, the transfer function, and low-pass behaviour. Building upon this demonstration, we outline a broader vision for an extensible library of circuit models, definitions, and verified results. Our long-term aim is to digitise, standardise, and verify circuit theory, enabling established knowledge to be systematically checked, shared, and reused across different circuit analyses. 
\end{abstract}

\section{Introduction}

Circuit theory provides the mathematical foundation for the analysis and design of electronic circuits across a wide range of domains, including analog, power, and radio-frequency (RF) electronics. Engineers use component models, Kirchhoff’s laws, and network equations to derive results for filters, amplifiers, and other circuits. These results depend on assumptions about component behaviour, circuit connections, and operating conditions. However, these assumptions are not always preserved when a formula is transferred from a textbook to a calculation or software implementation. Making such dependencies explicit is essential for ensuring that circuit analyses can be reliably checked, reproduced, and reused.

Interactive theorem provers (ITPs) provide a means of representing mathematical results alongside their assumptions and proofs. In an ITP, users define mathematical objects, state propositions, and construct proofs that are checked by the computer. Each result must follow logically from the definitions, assumptions, and previously established results on which it depends. This enables the verification of complex mathematical proofs from fundamental axioms \cite{buzzard2025flt}. The approach can reveal missing assumptions and logical errors that may be overlooked in informal derivations. However, formal verification only establishes mathematical correctness within a specified model. The validity of the model itself and its fidelity to physical circuit behaviour remain matters of engineering judgement.

Recent efforts to formalise scientific knowledge have demonstrated the potential of ITPs in fields beyond mathematics, including chemistry \cite{feyzishendi2025formalverificationisothermalchemical, 10.1039/d3dd00077j, ugwuanyi2026leanbetformallyverifiedsurfacearea} and physics \cite{toobysmith2025digitalizingwickstheorem, lu2017special, khanafshar2014optical, siddique2015optics}. Similar opportunities exist in circuit theory and analog electronics, where many seemingly distinct circuit analyses rely on common component laws, conservation principles, and mathematical operations. A shared formal foundation could allow new analyses to build on previously verified results while making the assumptions required for their application explicit.

In this work, we propose a framework based on Lean 4 \cite{de2015lean} that brings circuit models, assumptions, and mathematical proofs together in a machine-readable and verifiable form. The broader aim is to \textbf{digitise}, \textbf{standardise}, and \textbf{verify} circuit theory. Digitisation makes circuit reasoning accessible to software beyond the conventional representation of equations in textbooks and academic papers. While these equations are already available in digital form, the logical relationships connecting their premises and conclusions typically remain implicit. Formalisation records these relationships explicitly, enabling software to trace results to their assumptions and reuse established theorems in subsequent derivations. 

Standardisation provides a common foundation for describing and analysing different circuits. Filters, amplifiers, and larger networks rely on shared concepts such as terminal voltage, branch current, and Kirchhoff’s laws. Defining these concepts consistently allows contributors to reuse established definitions and distinguish precisely between different circuit models. For example, loaded and unloaded RC filters share the same fundamental component laws but differ in their network equations. 

Verification establishes that a circuit result follows logically from its stated model and assumptions. Lean 4 checks each step of a proof, including the conditions under which previously established theorems are applied. A theorem concerning gain or bandwidth can therefore establish a result for all parameter values satisfying its assumptions, without requiring a separate numerical evaluation for each case. Bobbin et al. demonstrated this approach in chemical physics by formalising established derivations and explicitly recording their premises \cite{10.1039/d3dd00077j}. In circuit analysis, such mathematical verification would complement simulation and experimental measurement, which remain essential for assessing the correspondence between theoretical models and physical circuits.

As an initial proof of concept, we consider an ideal, unloaded RC low-pass filter operating in sinusoidal steady state. We formalise its governing equations and derive its transfer function in Lean 4, demonstrating how circuit assumptions, mathematical definitions, and verified results can be represented within a common framework. Although elementary, this example illustrates the central methodology and provides a foundation for investigating the reuse of formalised circuit knowledge. For instance, introducing an output load modifies the network equations while preserving the underlying component laws, offering a natural opportunity to examine which definitions and theorems can be reused.

Beyond this initial demonstration, we envisage the development of a reusable library of circuit-theoretic definitions, models, and theorems. An important next step is to establish general representations of circuit connectivity, component behaviour, and network equations, allowing individual circuit analyses to be constructed from common building blocks rather than formalised independently. Such foundations could subsequently support more advanced topics, including dynamic circuit analysis, frequency-domain methods, and active-device models. These extensions introduce additional mathematical and modelling challenges, but also provide opportunities to demonstrate the scalability and practical value of the proposed approach.

This paper presents an initial step towards a broader programme of formalising circuit theory in Lean 4. Rather than treating individual circuit derivations as isolated verification tasks, we aim to develop a systematic approach in which circuit knowledge can be accumulated, independently checked, and reused. Ultimately, we envisage an extensible, community-developed library that serves as a common formal foundation for research and education in circuit theory and analog electronics.

\section{Context and Related Work}

An interactive theorem prover (ITP) is a software system that allows users to express mathematical definitions, state theorems, and construct proofs in a formal language \cite{toobysmith2025perspective}. Unlike conventional mathematical writing, in which the correctness of a derivation is assessed by human readers, an ITP checks whether each conclusion follows logically from the definitions, assumptions, and previously established results. Examples of ITPs include Lean 4, Rocq (formerly Coq) \cite{barras1997coq}, and Isabelle \cite{nipkow2002isabelle}. In this work, we adopt Lean 4, a functional programming language and interactive theorem prover supported by an extensive mathematical library known as Mathlib \cite{mathlib2020lean} and an enthusiastic online community \cite{leanprover_zulip}.

Lean 4 is based on dependent type theory, in which mathematical objects are represented as terms belonging to specified types \cite{10.1039/d3dd00077j}. For example, natural numbers, real numbers, and complex numbers are represented by different types $\mathbb{N}$, $\mathbb{R}$, and $\mathbb{C}$, respectively. Mathematical propositions are also represented as types, and proving a proposition amounts to constructing a term of the corresponding type. This correspondence allows Lean 4 to express mathematical statements and their proofs within a common logical framework.

In practice, users construct proofs interactively using \textit{tactics}, which apply logical rules and previously established theorems to transform a proof goal into simpler sub-goals. Lean 4 provides immediate feedback on the remaining goals and assumptions. Common tactics include \textbf{rw} for rewriting expressions using equalities, \textbf{simp} for simplifying expressions, and \textbf{ring} for proving polynomial identities. Although tactics automate many routine steps, the resulting proofs are ultimately checked by Lean 4's small trusted kernel, independently of the procedures used to construct them.

An important feature of Lean 4 is its ability to organise mathematical knowledge into reusable definitions and theorems. Through Mathlib, users can build upon existing formalised results in algebra, calculus, linear algebra, and other mathematical disciplines without reproducing their proofs. The use of Mathlib also ensures the correctness of definitions and allows changes in the broader ecosystem to be readily propagated to this project and downstream dependencies. This distinguishes Lean 4 from conventional computer algebra systems \cite{mathworks2023symbolic, wolfram1991mathematica}, which primarily focus on symbolic manipulation and numerical computation rather than constructing machine-checked proofs.

For circuit theory, these capabilities provide a means of representing component laws, network equations, and operating assumptions as formal mathematical objects. Circuit properties can then be derived from these definitions, with Lean 4 checking the logical validity of each derivation. Importantly, such verification establishes correctness relative to the specified mathematical model, rather than independently validating the model's correspondence to physical hardware. 

Note that this project is not currently based on Physlib\footnote{One of the authors of this paper (Zhi Kai Pong) is a maintainer of Physlib.} \cite{physlean}. This is in part due to the fact that electromagnetic theory is not yet sufficiently formalised in Physlib. We envision that this project could eventually be connected to Physlib, allowing the definitions of circuit quantities such as capacitance and inductance to be grounded more firmly in their underlying physical principles, for example, through the models of parallel-plate capacitors and solenoids, respectively.


\section{Formalisation of an RC Low-Pass Filter}
\label{sec:rc_formalisation}

To illustrate the proposed approach, we formalise a first-order RC low-pass filter in Lean 4. Rather than introducing its transfer function as an assumed expression, the implementation begins with impedance and component models, describes how two components are connected, and derives circuit-level results from these definitions. The example is restricted to an ideal, unloaded two-element series circuit in sinusoidal steady state, with complex-valued voltage and current phasors. Note that for the current scope of the project, we assume results of Linear Time-Invariant (LTI) systems, but do not currently plan to formalise advanced mathematical results from it e.g. Laplace transforms, Lyapunov stability, Nyquist criterion etc. However, we would like to eventually connect to other formally verified mathematics, such as Mathlib or standalone libraries e.g. \cite{doll2026foundations_arxiv}. The selected listings below present the principal definitions and theorem statements; proof bodies are omitted for brevity. The complete Lean 4 source code contains their machine-checked proofs.

\subsection{Representing Component Impedance}

A subtlety arises when modelling the impedance of a capacitor at zero frequency. For a non-zero capacitance and angular frequency, its phasor impedance is $Z_C=1/(j\omega C)$. At $\omega=0$, however, the ideal capacitor behaves as an open circuit. Simply evaluating this expression in Lean 4 would not capture that case: division is defined for all complex numbers, with $1/0=0$. Moreover, the usual form of Ohm's law, $V=IZ$, cannot directly accommodate an infinite impedance.

We therefore distinguish finite impedance from an open circuit using the inductive type in Listing~\ref{lst:impedance}. The associated constitutive relation imposes $V=IZ$ for a finite impedance, but $I=0$ for an open circuit. In the latter case, the voltage is not constrained by that constitutive relation alone.

\begin{lstlisting}[
    caption={Impedance and its constitutive relation},
    label={lst:impedance}
]
inductive Impedance where
  | finite (Z : ℂ)
  | openCircuit

def Impedance.ohmsLaw (Z : Impedance) (V I : ℂ) : Prop :=
  match Z with
  | .finite Z    => V = I * Z
  | .openCircuit => I = 0
\end{lstlisting}

The component models in Listing~\ref{lst:passive-impedance} use non-negative real-valued parameters. A resistor is assigned its real-valued impedance, while a capacitor is represented as an open circuit when either its capacitance or the angular frequency is zero. This case distinction makes the DC limit explicit rather than relying on an informal interpretation of a singular expression.

\begin{lstlisting}[
    caption={Resistor and capacitor impedance models},
    label={lst:passive-impedance}
]
def resistor (R : ℝ≥0) : Impedance :=
  .finite R

noncomputable def capacitor (C : ℝ≥0) (ω : ℝ≥0) : Impedance :=
  if ω = 0 ∨ C = 0 then .openCircuit else .finite (1 / (j * ω * C))
\end{lstlisting}

\subsection{From Components to Circuit Constraints}

A \texttt{Component} associates an impedance with a complex voltage and current, together with a proof that these quantities satisfy the corresponding constitutive relation. The circuit representation then combines two such components in series. As shown in Listing~\ref{lst:series-circuit}, its fields record the equality of branch currents, the source and intermediate-node potentials, and the voltage drops across the two components. All node potentials are measured relative to a reference ground node, which defaults to zero.

\begin{lstlisting}[
    caption={Component and two-element series-circuit definitions},
    label={lst:series-circuit}
]
structure Component (Z : Impedance) where
  voltage : ℂ
  current : ℂ
  law : Z.ohmsLaw voltage current

structure TwoElementSeriesCircuit (Z1 Z2 : Impedance) where
  c1 : Component Z1
  c2 : Component Z2
  current_equal : c1.current = c2.current
  src : ℂ
  node : ℂ
  gnd : ℂ := 0
  v1 : c1.voltage = (src - gnd) - (node - gnd)
  v2 : c2.voltage = node - gnd
\end{lstlisting}

These constraints encode the connectivity of this particular network. The implementation also proves Kirchhoff's current and voltage laws for the two-element circuit: $I_1-I_2=0$ and $V_1+V_2-(V_{\mathrm{src}}-V_{\mathrm{gnd}})=0$. Here, these results follow from the current-equality and branch-voltage fields of the structure; they do not yet constitute a general-purpose representation of arbitrary circuit topology. Nevertheless, separating component behaviour from connection constraints provides a useful starting point for reusing the same component models in other configurations.

\subsection{Reusable Voltage-Division Results}

The transfer function of the two-element circuit is defined as the ratio of the output-node phasor voltage to the input phasor voltage, both referred to ground. For finite impedances $z_1$ and $z_2$, Lean 4 establishes the familiar voltage-divider identity
\begin{equation}
H=\frac{V_{\mathrm{out}}}{V_{\mathrm{in}}}=\frac{z_2}{z_1+z_2},
\label{eq:general-divider}
\end{equation}
subject to a non-zero input voltage and a non-zero sum of impedances. Listing~\ref{lst:divider} shows the corresponding declarations. A separate theorem covers the case where the second impedance is an open circuit: provided the first impedance is finite and the input voltage is non-zero, the transfer function equals unity. The source also proves that an open first component and finite second component give zero transfer.

\begin{lstlisting}[
    caption={Transfer-function definition and selected voltage-divider theorem statements; proofs omitted},
    label={lst:divider}
]
noncomputable def tf (c : TwoElementSeriesCircuit Z1 Z2) : ℂ :=
  (c.node - c.gnd) / (c.src - c.gnd)

theorem tf_eq_of_finite (c : TwoElementSeriesCircuit Z1 Z2)
    {z1 z2 : ℂ} (hZ1 : Z1 = .finite z1) (hZ2 : Z2 = .finite z2)
    (hsrc : c.src - c.gnd ≠ 0) (hsum : z1 + z2 ≠ 0) :
    c.tf = z2 / (z1 + z2)

theorem tf_eq_of_open_right (c : TwoElementSeriesCircuit Z1 Z2)
    {z1 : ℂ} (hZ1 : Z1 = .finite z1) (hZ2 : Z2 = .openCircuit)
    (hsrc : c.src - c.gnd ≠ 0) :
    c.tf = 1
\end{lstlisting}

The significance of these results is their separation from any particular choice of resistor or capacitor. Once established for the two-element circuit structure, they can be invoked for different constituent impedances without repeating the voltage-division derivation.

\subsection{RC Transfer Function and Low-Pass Behaviour}

The RC low-pass filter is obtained by assigning a resistor to the first series element and a capacitor to the second, taking the intermediate node as the output. Listing~\ref{lst:rc-transfer} gives this specialisation and the corresponding closed-form transfer function,
\begin{equation}
H(j\omega)=\frac{1}{1+j\omega RC}.
\label{eq:rc-transfer}
\end{equation}
The theorem \texttt{RC.tf\_eq} establishes that the transfer ratio computed from the circuit constraints equals this expression for all non-negative $R$, $C$, and $\omega$, assuming a non-zero source phasor. Its proof uses the finite-impedance divider theorem when $\omega\ne0$ and $C\ne0$, and the open-circuit theorem otherwise. Thus, the result includes DC without evaluating the capacitor's impedance through division by zero.

\begin{lstlisting}[
    caption={RC specialisation and transfer-function theorem statement; proof omitted},
    label={lst:rc-transfer}
]
abbrev RC.Circuit (R C ω : ℝ≥0) : Type :=
  TwoElementSeriesCircuit (resistor R) (capacitor C ω)

noncomputable def RC.tf (R C : ℝ≥0) : ℝ≥0 → ℂ :=
  fun ω => 1 / (1 + j * ω * R * C)

theorem RC.tf_eq {R C ω : ℝ≥0} (c : RC.Circuit R C ω) (hsrc : c.src - c.gnd ≠ 0) :
    c.tf = RC.tf R C ω
\end{lstlisting}

Finally, the implementation formalises one precise meaning of \emph{low-pass}: the complex frequency response approaches unity as the nonnegative angular frequency tends to zero, and approaches zero as frequency tends to infinity. Under strictly positive resistance and capacitance, Lean 4 proves that the derived RC transfer function satisfies both limits (Listing~\ref{lst:lowpass}). A further theorem transfers the property to a family of circuit instances with nonzero source voltage at every frequency. This definition captures the two limiting behaviours, rather than asserting, for example, monotonic magnitude or a particular cutoff frequency.

\begin{lstlisting}[
    caption={Low-pass definition and theorem statements; proofs omitted},
    label={lst:lowpass}
]
def IsLowpass (H : ℝ≥0 → ℂ) : Prop :=
  Tendsto H (�� 0) (�� 1) ∧ Tendsto H atTop (�� 0)

theorem RC.isLowpass {R C : ℝ≥0} (hR : 0 < R) (hC : 0 < C) :
    IsLowpass (RC.tf R C)
\end{lstlisting}

This example illustrates a progression from mathematical representations of components, through explicit circuit constraints, to reusable theorems about transfer functions and frequency response. Its scope remains deliberately limited: the implementation concerns an ideal two-element series network rather than general nodal analysis, loading, tolerances, or nonideal devices. Nonetheless, it provides a concrete, machine-checked starting point for developing more general representations of connectivity and for building circuit analyses from shared definitions and results.

\section{Conclusion and Outlook}

This paper presents an initial step towards a formalised framework for circuit theory and analog electronics using Lean 4. Through the example of an RC low-pass filter, we demonstrate how component models, circuit constraints, and mathematical results can be represented and verified within a common formal environment. The implementation establishes results for voltage division, the RC transfer function, and its low-pass behaviour, illustrating how circuit-level theorems can be constructed from more fundamental definitions and assumptions.

Looking ahead, our ambition extends beyond the verification of individual circuit equations. We envisage the development of a comprehensive and reusable library of circuit theory, in which component models, network laws, and established analytical results form an interconnected body of machine-checked knowledge. A key priority is to develop general abstractions for circuit topology and connectivity, enabling the analysis of increasingly complex networks without repeatedly defining the same underlying principles. Building upon these foundations, future developments could encompass transient analysis, frequency-domain methods, active-device models, and even non-linear circuits. Particular attention will be given to the reuse of definitions and theorems across different circuits, as this is essential for transforming isolated formalisation exercises into a scalable framework.

In the longer term, we hope that this effort will contribute to a broader ecosystem of formalised engineering knowledge, building upon existing mathematical foundations in Mathlib and potentially connecting with initiatives such as Physlib. Such integration could enable circuit models to be related more closely to their underlying physical principles, while maintaining the abstractions required for practical engineering analysis. Beyond verifying established results, a sufficiently mature library could support the systematic exploration of circuit properties, identification of overlooked assumptions, and development of computer-assisted reasoning tools for circuit research and education.

The present implementation remains a work in progress. We intend to release the Lean 4 source code as an open-source repository on GitHub in the near future, together with documentation to facilitate its use and extension. We welcome contributions from the wider circuits, mathematics, and formal methods communities, particularly towards improving existing definitions, expanding circuit coverage, and establishing reusable formalisation conventions. Ultimately, our goal is to encourage a collaborative effort towards a shared, machine-checkable foundation for circuit theory, in which established knowledge can be progressively formalised, refined, and extended.
    
 {\small
\bibliographystyle{IEEEtran}
\bibliography{egbib}
}


\end{document}